# Risk of Interruption of Doctoral Studies and Mental Health in PhD Students

**Sara M. González-Betancor [1,\*] and Pablo Dorta-González [2]**

[1] Departamento de Métodos Cuantitativos en Economía y Gestión, Facultad de Economía Empresa y Turismo, Universidad de Las Palmas de Gran Canaria, 35017 Las Palmas de Gran Canaria, Spain
[2] Department of Quantitative Methods in Economics and Management, and Institute TiDES, University of Las Palmas de Gran Canaria, 35017 Las Palmas de Gran Canaria, Spain; pablo.dorta@ulpgc.es
\* Correspondence: sara.gonzalez@ulpgc.es

**Abstract:** PhD students report a higher prevalence of mental illness symptoms than highly educated individuals in the general population. This situation presents a serious problem for universities. Thus, the knowledge about this phenomenon is of great importance in decision-making. In this paper we use the *Nature* PhD survey 2019 and estimate several binomial logistic regression models to analyze the risk of interrupting doctoral studies. This risk is measured through the desire of change in either the supervisor or the area of expertise, or the wish of not pursue a PhD. Among the explanatory factors, we focus on the influence of anxiety/depression, discrimination, and bullying. As control variables we use demographic characteristics and others related with the doctoral program. Insufficient contact time with supervisors, and exceeding time spent studying -crossing the 50-h week barrier-, are risk factors of PhD studies interruption, but the most decisive risk factor is poor mental health. Universities should therefore foster an environment of well-being, which allows the development of autonomy and resilience of their PhD students or, when necessary, which fosters the development of conflict resolution skills.

## 1. Introduction

Recent studies [1–3] have brought to light again the mental health problems suffered by students during their doctoral studies (hereafter referred to as PhD students). Although the interest about the well-being of students in higher education is not new [4–6], most authors have restricted their analyses to undergraduate and master studies. However, a PhD has its own characteristics, which make it worthy of a separate analysis from other postgraduate studies.

PhD students often complain of social isolation, loss of motivation, and communication difficulties with the supervisor [7–9]. The study of these factors has focused mainly on explaining the success or failure in the studies [10,11] as well as the assessment of mental illness symptoms [9,12–14].

Rates of anxiety and depression are high among the PhD students [14]. Moreover, PhD students report a higher prevalence of mental illness symptoms than highly educated individuals in the general population and other higher education students [14].

The six-factor model of psychological well-being provides a theoretical framework to understand the doctoral degree context [2]. Some authors analyze the social support [15,16] and autonomy [17]. Some other studies focus on conceptualizing and measuring the well-being of PhD students [18]. And others show that levels of well-being are correlated to progress in doctoral studies, professional development, and scientific productivity [10,19].

Low levels of well-being in PhD students suppose serious problems for the universities. PhD students make a significant contribution to the overall research output from universities [20,21], and low levels of well-being reduce the quality and quantity of the research outputs [22].

Universities are responsible for maintaining an environment that supports PhD students' well-being. In this respect, the knowledge of the mental health situation of the PhD students and its influence in the risk of interruption of doctoral studies is of great importance for decision-making. Poor mental health, which is linked to low levels of well-being, is associated with an increased risk of interruption in the studies [23–26].

In this paper we also analyze the risk of interruption of doctoral studies, but using a different methodology from that of previous studies, as well as an updated database. Specifically, we estimate several binomial logistic regression models using the data of the *Nature* PhD Survey 2019. The risk of interruption is measured through the desire of change in either the area of expertise or the supervisor, or the wish of not pursue a PhD. Among the explanatory factors, we focus on the influence of anxiety/depression, discrimination, and bullying. As control we use demographic variables and those other related with the doctoral program.

## 2. Materials and Methods

### 2.1. The Data

The *Nature* team have run a biennial PhD Career survey since 2011. In this research we use the last wave of the survey, namely *Nature* PhD survey 2019. This online survey was developed in collaboration with *Nature* and sent to their database and subscribers via different channels. In order to boost response in specific regions which have been previously under-represented, the survey was translated into four languages (Mandarin Chinese, Portuguese, Spanish and French) in addition to English. The survey was live for approximately six weeks during the months of May and June of 2019. The final usable sample, after removing poor quality responses and missing data, reaches 6320 records [27].

The survey included up to 56 questions, and we focus our research on the following one: "What would you do differently right now if you were starting your PhD program?"

The risk of interruption, thus, is measured through the desire of change in either the area of expertise or the supervisor, or the wish of not pursue a PhD. We consider that a desire to change (not to change) shows a dissatisfaction (satisfaction) of the PhD student with the PhD studies, and that dissatisfaction can materialize later in an interruption of studies.

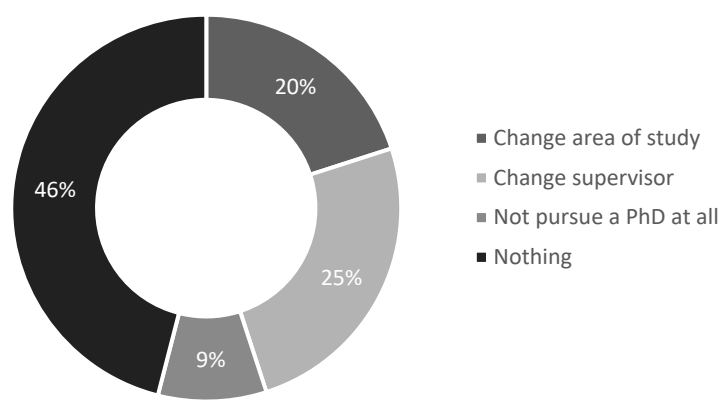

**Figure 1.** What would you do differently right now if you were starting your PhD program? Distribution of respondents (n = 6320).

As can be seeing in Figure 1, more than half of the respondents to the *Nature* PhD Survey 2019 would radically change the beginning of their PhD (changing area, supervisor or directly not pursuing a PhD at all). Less than half are satisfied with their initial choice, as they would not change anything.

*2.2. Method*

The variable under study discriminates between more options than the ones we are interested in. That is why we chose to combine the different alternatives into only two options: (a) not changing anything, (b) changing something (area, supervisor or not doing a PhD at all). In this way we transformed the variable object of study into a dichotomous one and estimated the probability of belonging to the group of students who would not change anything, as opposed to the probability of belonging to the group of students who would change something.

To estimate this probability, we chose to carry out the maximum likelihood estimation of a binomial logit model [28], in which we introduced a set of explanatory control variables -related to demographic characteristics of the students, as well as to characteristics of the doctoral program itself-, together with the variables under study -related to mental health characteristics-. In this way we can see which personal, doctoral program or mental health characteristics increase the probability of belonging to one group or another.

As there are some missing values among the explanatory variables, the sample is reduced from 6320 to 5911 PhD students. Nevertheless, we decided not to use multiple assignment techniques, though the loss of sample size, due to the uncertainty of the quality of the assignment of such values [29]. The estimation of the logit model is made for the entire sample without missing values (5911), and repeated by splitting the sample in two: students who had no mental health problems (2908) and those who had mental health problems -anxiety or depression, or suffered discrimination or bullying- (3003). In this way we try to identify different patterns of behaviour in terms of the explanatory variables and the probability of belonging to each of the groups (those who would not change anything vs. those who would).

Finally, we repeat the estimations with other subsamples. In this case, differentiating the students with some problem according to the problem typology. In this way we try to see if the explanatory variables have a different influence on the group of those who have suffered anxiety/depression (2164), those who have suffered discrimination (1252) and those who have suffered bullying (1280).

All the estimated models try to analyse which variables increase the probability of belonging to the group of those who would not modify their choice of PhD, since the generated dichotomous variable $Y_i$ takes the value 1 when students report that choice and the value 0 when they report that they would modify something (area, supervisor or not doing a PhD at all). Equation (1) shows this probability as a function of the k-explanatory variables ($X_{ki}$), whose summary statistics are shown in the Appendix (Table A1):

$$E(Y_i/X_{ki}) = Pr(Y_i = 1/X_{ki}) \tag{1}$$

The expected values of the dependent variable $Y_i$, using the logistic function, can be transformed in the logistic model (2):

$$logit[Pr(Y_i = 1/X_{ki})] = logit\ P_i = ln\left(\frac{P_i}{1-P_i}\right) = \beta_0 + \sum_k \beta_k X_{ki} \tag{2}$$

where $P_i$ is the probability of not modifying anything related to the PhD choice, $\beta_0$ is the mean probability for the reference student, and $\beta_k$ the coefficients of the k-explanatory variables. Thus, the probability of not modifying anything related to their PhD choice can be expressed as (3):

$$Pr(Y_i = 1/X_{ki}) = \frac{e^{\beta_0 + \sum_k \beta_k X_{ki}}}{1 + e^{\beta_0 + \sum_k \beta_k X_{ki}}} \tag{3}$$

Equation (3) has been estimated with the abovementioned sample and subsamples. All estimations and post-estimations have been performed with the StataSE 15 statistical package using the logit command [30].

## 3. Results

Equation (3) is estimated for the whole sample (Logit 1), for the subsample of students with no mental health problems (Logit 2), and finally for the subsample of students with mental health problems (Logit 3). Results are shown on Table 1, where it is easy to interpret which variables increase the probability of not wanting to change anything related to the PhD, and which variables decrease this probability or, looked at another way, which variables increase the probability of wanting to change something.

As all variables except age are categorical, Table 1 shows the estimations in the form of odd-ratios. When the value of the odd-ratio is greater (lower) than one, the variable increases (decreases) the probability of not wanting to change anything. Therefore, students with the characteristics of those variables that show an odd-ratio greater than one would maintain their choice of PhD without any modification, and those with characteristics that show an odd-ratio lower than one would change area or supervisor or, directly, would not pursue a PhD.

In addition to the odd-ratios, Table 1 shows the 95% confidence intervals, so that, for those variables that are statistically significant, it can be checked whether their confidence intervals include the value 1, in which case it cannot be stated whether this variable increases or decreases the corresponding probability.

The three estimated models show good results in terms of goodness of fit, since they manage to correctly classify more than 70% of the sample and have a pseudo coefficient of determination of 0.20, 0.13, and 0.19, respectively.

Out of all demographic variables, only age is significant in the Logit 1 and Logit 3 estimates. Their odd-ratio is less than 1, so as the student's age increases the probability of wanting to change also increases. All other variables, such as gender, race, region of birth, having childcare or eldercare responsibilities, or being working while studying, do not have a statistically significant influence on this probability.

Regarding the characteristics of the doctoral program, the estimations show that the most decisive factor is the students' previous expectations. When the doctoral program does not meet the original expectations, the student's probability of not wanting to change anything decreases very sharply, and when it exceeds those expectations the opposite occurs (the probability of not wanting to change anything increases greatly).

The next variable with the highest odd-ratio value is the one that refers to the weekly contact time with the supervisor. As this weekly contact time increases, so does the probability of not wanting to change anything in all three models (Logit 1, Logit 2, and Logit 3).

On the other hand, the weekly time spent on the PhD program is only significant—but lower than 1—when it exceeds the fifty hour per week barrier. In that case, the probability of not wanting to change anything decreases (or the probability of wanting to change something increases). However, in the group of those who have had some kind of problem (Logit 3), this variable is not statistically significant.

Finally, regarding Table 1, the three variables related to mental health are statistically significant in the two models in which they intervene (Logit 1 and Logit 3), showing that students who have suffered anxiety or depression, discrimination or bullying, are the least likely to want to keep everything unchanged. Especially in the case of those who have experienced bullying. The Logit 3 estimation was made to the subsample of students who had stated that they had suffered some kind of problem, without distinguishing the type of problem. In order to check if there are differences in the influence of the variables according to the problem suffered by the students, we chose to estimate three new binomial logistic regressions (Logit 4, Logit 5, Logit 6) and show their results in. The models have been estimated with the same explanatory variables as those in Table 1, but here we show only the variables of interest: those related to the doctoral program and to mental health.

Table 1. Binomial logistic regression (endogenous variable: Do nothing vs. Change area/supervisor/not pursue a PhD).

| | Logit 1 | | | Logit 2 | | | Logit 3 | | |
|---|---|---|---|---|---|---|---|---|---|
| | Whole Sample | | | PhD Students with No Problems | | | PhD Students with Problems | | |
| | O.R. | Sig. | 95% CI | O.R. | Sig. | 95% CI | O.R. | Sig. | 95% CI |
| Constant | 2.9743 | *** | (2.0036–4.4594) | 3.2220 | *** | (1.9332–5.4161) | 2.7732 | *** | (1.4188–5.4418) |
| **DEMOGRAPHIC VARIABLES** | | | | | | | | | |
| Age | 0.9844 | *** | (0.9732–0.9958) | 0.9863 | * | (0.9720–1.0009) | 0.9812 | ** | (0.9630–0.9998) |
| **Gender** | | | | | | | | | |
| Male (including trans male) | Ref. | | | Ref. | | | Ref. | | |
| Female (including trans female) | 0.9793 | | (0.8654–1.1082) | 1.0239 | | (0.8649–1.2120) | 0.9175 | | (0.7631–1.1033) |
| Gender queer/Non binary | 0.7550 | | (0.2987–1.9092) | 0.2060 | * | (0.0378–1.1324) | 1.5311 | | (0.4909–4.7721) |
| **Race** | | | | | | | | | |
| Caucasian | Ref. | | | Ref. | | | Ref. | | |
| Latino-Hispanic | 0.8650 | | (0.6882–1.0882) | 0.8668 | | (0.6176–1.2158) | 0.8923 | | (0.6514–1.2232) |
| Middle Eastern | 0.8445 | | (0.5996–1.1899) | 0.8017 | | (0.4832–1.3304) | 0.8781 | | (0.5496–1.4027) |
| African | 1.4148 | * | (0.9387–2.1339) | 1.5189 | | (0.8325–2.7690) | 1.3364 | | (0.7504–2.3786) |
| Caribbean | 0.8378 | | (0.3631–1.9347) | 1.3689 | | (0.3871–4.8407) | 0.5031 | | (0.1457–1.7366) |
| South Asian | 1.0177 | | (0.8100–1.2785) | 0.9807 | | (0.7161–1.3430) | 1.0960 | | (0.7833–1.5337) |
| East Asian | 0.9254 | | (0.7507–1.1408) | 0.8994 | | (0.6693–1.2075) | 0.9861 | | (0.7313–1.3296) |
| Pacific Islander | 1.1411 | | (0.5810–2.2391) | 1.3258 | | (0.4738–3.7091) | 0.9747 | | (0.3866–2.4581) |
| American Indian | 1.0515 | | (0.2940–3.7609) | 1.3073 | | (0.0797–21.4242) | 0.9402 | | (0.2200–4.0171) |
| Mixed ethnicity | 0.8454 | | (0.5838–1.2232) | 1.0988 | | (0.6561–1.8400) | 0.6440 | | (0.3635–1.1420) |
| **Region** | | | | | | | | | |
| Europe | Ref. | | | Ref. | | | Ref. | | |
| Africa | 0.8106 | | (0.4939–1.3311) | 0.9458 | | (0.4663–1.9184) | 0.6771 | | (0.3243–1.4124) |
| Asia (including Middle East) | 0.9986 | | (0.7995–1.2473) | 0.8825 | | (0.6462–1.2045) | 1.1677 | | (0.8461–1.6109) |
| Australasia | 1.0457 | | (0.7413–1.4751) | 0.9629 | | (0.6040–1.5351) | 1.1653 | | (0.6966–1.9511) |
| North or Central America | 0.9688 | | (0.8287–1.1326) | 0.8319 | | (0.6683–1.0366) | 1.1343 | | (0.9034–1.4230) |
| South America | 0.7953 | | (0.5618–1.1263) | 0.7393 | | (0.4518–1.2089) | 0.8720 | | (0.5271–1.4418) |
| **Caring** | | | | | | | | | |
| No | Ref. | | | Ref. | | | Ref. | | |
| Yes | 1.0356 | | (0.8777–1.2219) | 0.9693 | | (0.7720–1.2169) | 1.1208 | | (0.8791–1.4292) |
| **Job** | | | | | | | | | |
| No | Ref. | | | Ref. | | | Ref. | | |
| Yes | 0.9381 | | (0.7975–1.1035) | 0.9194 | | (0.7343–1.1512) | 0.9564 | | (0.7544–1.2124) |

**Table 1.** *Cont.*

| | | | | | | |
|---|---|---|---|---|---|---|
| **VARIABLES RELATED WITH THE DOCTORAL PROGRAM** | | | | | | |
| *Same country as grown* | | | | | | |
| No | Ref. | | Ref. | | Ref. | |
| Yes | 1.0292 | (0.8945–1.1843) | 0.9532 | (0.7871–1.1544) | 1.1174 | (0.9061–1.3786) |
| *PhD related to original expectation* | | | | | | |
| Meets original expectations | Ref. | | Ref. | | Ref. | |
| Does not meet original expectations | 0.1670 *** | (0.1452–0.1924) | 0.1827 *** | (0.1501–0.2242) | 0.1526 *** | (0.1251–0.1862) |
| Exceeds original expectations | 2.0668 *** | (1.6840–2.5349) | 2.4327 *** | (1.8231–3.2446) | 1.7212 *** | (1.2752–2.3210) |
| *Hours/week spent on PhD program* | | | | | | |
| Less than 40 h | Ref. | | Ref. | | Ref. | |
| 41–50 h | 0.9093 | (0.7689–1.0752) | 0.8479 | (0.6782–1.0606) | 0.9805 | (0.7591–1.2664) |
| More than 50 h | 0.8212 ** | (0.7041–0.9576) | 0.8017 ** | (0.6489–0.9896) | 0.8420 | (0.6704–1.0569) |
| *Hours/week contact with supervisor* | | | | | | |
| Less than an hour | Ref. | | Ref. | | Ref. | |
| Between one and three hours | 1.4859 *** | (1.3042–1.6942) | 1.4120 *** | (1.1790–1.6902) | 1.5888 *** | (1.3111–1.9258) |
| More than three hours | 1.7212 *** | (1.4113–2.0971) | 1.8739 *** | (1.4294–2.4565) | 1.5621 *** | (1.1611–2.1030) |
| Other, please specify | 0.9103 | (0.6658–1.2445) | 0.6744 * | (0.4254–1.0696) | 1.2141 | (0.7983–1.8455) |
| **MENTAL HEALTH VARIABLES** | | | | | | |
| *Seek help for anxiety or depression* | | | | | | |
| No | Ref. | | | | Ref. | |
| Yes | 0.5993 *** | (0.5260–0.6823) | | | 0.6005 *** | (0.4810–0.7499) |
| *Discrimination* | | | | | | |
| No | Ref. | | | | Ref. | |
| Yes | 0.7796 *** | (0.6551–0.9278) | | | 0.7906 ** | (0.6466–0.9660) |
| *Bullying* | | | | | | |
| No | Ref. | | | | Ref. | |
| Yes | 0.5347 *** | (0.4491–0.6367) | | | 0.5358 *** | (0.4397–0.6536) |
| Chi2 | 1601 | | 524 | | 720 | |
| Pearson chi2 | 4088.5 | | 1715.16 | | 2379.69 | |
| DF | 3931 | | 1613 | | 2290 | |
| Prob > chi2 | 0.0392 | | 0.0381 | | 0.0937 | |
| Correctly classified | 71.66% | | 70.43% | | 73.83% | |
| Pseudo R2 | 0.1965 | | 0.1328 | | 0.1889 | |
| N | 5911 | | 2908 | | 3003 | |

Concerning the variables related to the doctoral program, it can be seen that expectations are still relevant. When the program does not meet the student's expectations, the probability of wanting to keep everything unchanged decreases in all three groups, especially among those who have suffered discrimination (although the ICs in the three models overlap, so the difference between them is probably not significant). However, for the subsample of those who have had anxiety/depression (Logit 4) the fact that the program exceeds the students' expectations is not relevant to the probability under study, unlike the other two subsamples (Logit 5 and Logit 6).

Regarding the contact time with the supervisor, there are again differences between subsamples. The subsample of those who have had anxiety or depression (Logit 4) presents odd-ratios for this variable almost identical to those of the total sample (Logit 1). However, the subsample of those who have suffered discrimination or bullying only value positively having a weekly contact with their supervisor of between one and three hours, but not more than three hours a week.

Finally, in the subsample of those who have suffered from anxiety or depression (Logit 4), having also suffered from bullying is a determining factor in reducing the probability of wanting to keep everything unchanged, while discrimination is not statistically significant enough to influence this probability. In the group of those who have suffered discrimination, both having had anxiety/depression, and having experienced bullying, decreases in the same proportion the probability of wanting to keep everything unchanged. However, in the group of those who have experienced bullying, neither anxiety/depression nor discrimination is statistically significant enough to influence this probability.

## 4. Discussion and Conclusions

Low levels of well-being in PhD students are a serious problem for universities. In decision-making, the awareness about the mental health situation of PhD students and its influence on the risk of interruption of doctoral studies is of great relevance.

In this respect we estimated several binomial logistic regression models in a large-scale survey of about six thousand PhD students. The risk of interruption of the studies was measured through the desire of changing area of expertise, or supervisor, or to regret having chosen to pursue a PhD. Among the explanatory factors, we focused on the influence of three mental health aspects: anxiety/depression, discrimination, and bullying. As control variables we used some demographic variables and some others related with the doctoral program itself.

Regarding the demographic variables, only age was statistically significant for the risk of interruption: as the student's age increases the risk of interruption also increases. All other demographic variables, such as gender, race, region of birth, having childcare or eldercare responsibilities, or being working while studying, do not have a statistically significant influence on the risk of interruption in the studies.

Regarding the characteristics of the doctoral program itself, the most decisive risk factor is not meeting the students' original expectations. When the doctoral program does not meet the original expectations, the risk of interruption increases considerably, and when it exceeds those expectations the opposite occurs. The second most decisive risk factor is insufficient contact time with the supervisor. As the weekly contact time decreases, the risk of interruption increases. On the other hand, the weekly time spent on the PhD program is only significant when it exceeds the fifty hour per week barrier. In that case, the risk of interruption increases. However, in the group of those who have had some mental health problem, this variable is not statistically significant, which leads us to think about the possibility that these students have a higher level of resilience or endurance and that, therefore, it is not relevant for them that their weekly time spent on the PhD program exceeds fifty hours.

**Table 2.** Binomial logistic regressions (endogenous variable: Do nothing vs. Change area/supervisor/not pursue a PhD)—Subsample: Students with mental health problems.

| | Logit 4 | | | Logit 5 | | | Logit 6 | | |
|---|---|---|---|---|---|---|---|---|---|
| | **Students with Anxiety/Depression** | | | **Students Who Suffer Discrimination** | | | **Students Who Suffer Bullying** | | |
| | O.R. | Sig. | 95% CI | O.R. | Sig. | 95% CI | O.R. | Sig. | 95% CI |
| **VARIABLES RELATED WITH THE DOCTORAL PROGRAM** | | | | | | | | | |
| *Same country as grown* | | | | | | | | | |
| No | Ref. | | | Ref. | | | Ref. | | |
| Yes | 1.1044 | | (0.8560–1.4249) | 1.1960 | | (0.8492–1.6836) | 1.1912 | | (0.8266–1.7154) |
| *PhD related to original expectation* | | | | | | | | | |
| Meets original expectations | Ref. | | | Ref. | | | Ref. | | |
| Does not meet original expectations | 0.1541 | *** | (0.1212–0.1954) | 0.1395 | *** | (0.1000–0.1925) | 0.1466 | *** | (0.1059–0.2044) |
| Exceeds original expectations | 1.3951 | * | (0.9850–1.9775) | 2.1858 | *** | (1.2612–3.7891) | 2.5831 | *** | (1.4801–4.5075) |
| *Hours/week spent on PhD program* | | | | | | | | | |
| Less than 40 h | Ref. | | | Ref. | | | | | |
| 41–50 h | 0.9709 | | (0.7194–1.3104) | 0.9879 | | (0.6219–1.5693) | 1.0645 | | (0.6626–1.7103) |
| More than 50 h | 0.8245 | | (0.6323–1.0756) | 0.8049 | | (0.5356–1.2109) | 0.9003 | | (0.5952–1.3615) |
| *Hours/week contact with supervisor* | | | | | | | | | |
| Less than an hour | Ref. | | | Ref. | | | Ref. | | |
| Between one and three hours | 1.5311 | *** | (1.2166–1.9254) | 1.9798 | *** | (1.4304–2.7413) | 1.5174 | ** | (1.0900–2.1117) |
| More than three hours | 1.7126 | *** | (1.2130–2.4174) | 1.6128 | * | (0.9482–2.7458) | 1.3840 | | (0.8251–2.3210) |
| Other, please specify | 1.0871 | | (0.6455–1.8306) | 0.8642 | | (0.4501–1.6593) | 0.8479 | | (0.4338–1.6585) |
| **MENTAL HEALTH VARIABLES** | | | | | | | | | |
| *Seek help for anxiety or depression* | | | | | | | | | |
| No | | | | Ref. | | | Ref. | | |
| Yes | | | | 0.6090 | *** | (0.4513–0.8223) | 0.4986 | | (0.3658–0.6793) |
| *Discrimination* | | | | | | | | | |
| No | Ref. | | | | | | Ref. | | |
| Yes | 0.8212 | | (0.6282–1.0736) | | | | 0.9490 | *** | (0.6926–1.3006) |
| *Bullying* | | | | | | | | | |
| No | Ref. | | | Ref. | | | | | |
| Yes | 0.4834 | *** | (0.3695–0.6328) | 0.6139 | *** | (0.4553–0.8283) | | | |
| N | 2164 | | | 1252 | | | 1280 | | |

Note: Each of these models include the same explanatory variables as those of Table 1, but we only show the variables we are interested at, for explanatory purposes.

Finally, the three variables related to mental health are statistically significant in all the models analyzed, showing that students who have suffered anxiety or depression, discrimination or bullying, are the more likely to interrupt the doctoral studies, especially in the case of those who have experienced bullying.

Distinguishing by subsamples according to the mental health problem, the initial expectations are still relevant. When the doctoral program does not meet the student's expectations, the risk of interruption increases. However, for the subsample of those who have had anxiety/depression the fact that the program exceeds the students' expectations is not relevant in reducing the risk of interruption, unlike the other two subsamples of those who have suffered discrimination or bullying.

Regarding the contact time with the supervisor, there are again differences between subsamples. The subsample of those who have had anxiety or depression presents values almost identical to those of the total sample. However, the subsample of those who have suffered discrimination or bullying only value positively having a weekly contact with their supervisor of between one and three hours, but not more than three hours a week. It is possible that, due to their previous poor interpersonal relationship experience, a contact of more than three hours per week with the supervisor may not be for them a sign of better quality of a doctoral program.

Finally, in the subsample of those who have suffered from anxiety or depression, having also suffered from bullying is a determining factor in the risk of interruption, while discrimination is not statistically significant. In the group of those who have suffered discrimination, both having had anxiety/depression, and having experienced bullying, increase in the same proportion the risk of interruption. However, in the group of those who have experienced bullying, neither anxiety/depression, nor discrimination is statistically significant enough to influence this risk of interruption.

Universities are in charge of maintaining an environment that supports PhD students' well-being. Suggestions for improving the well-being of doctoral students include three levels of intervention. Firstly, actions in relation to the development of student resilience and autonomy [31,32]. Secondly, actions to promote positive relationships between students and their supervisors, with clear guidelines for both student and supervisor expectations. And finally, training to both student and supervisors about conflict resolution and relationship boundaries [16,17,33].

The contribution of this paper to the literature is mainly methodological. We analyzed the risk of interruption of doctoral studies using a different methodology from that of previous studies in the literature, as well as an updated database. We estimated several binomial logistic regression models using data from a 2019 survey. We measured the risk of interruption through the desire of change in either the area of expertise or the supervisor, or the wish of not to pursue a PhD. Among the explanatory factors we focused on the influence of anxiety/depression, discrimination, and bullying, while using other demographic variables, as well as variables related to the doctoral program, as control ones.

The main limitation of this study lies in the fact that we can only observe the 'risk' of abandoning doctoral studies, since the survey was carried out among students who were actually doing their doctorate and therefore none of them had abandoned it. In fact, the desire to change the area of study or supervisor does not necessarily imply abandonment of studies. Thus, we cannot speak of a 'dropout rate', but only of a risk of abandonment. However, the higher the risk, the greater the likelihood of dropout.

Future research could extend the study to a sample that includes both types of students, those who have completed their doctoral studies and those who have left their studies uncompleted, in order to explore what the main dropout factors may have been.

**Funding:** This work has been partly supported by the Ministerio de Economía, Industria y Competitividad under Research Project ECO2017-88883-R and the FEDER funding under Research Project UMA18FEDERJA024.

**Conflicts of Interest:** The authors declare no conflict of interest.

## Appendix A

**Table A1.** Descriptive statistics of the explanatory variables.

| Variables | Categories | Do Nothing | Change |
|---|---|---|---|
| Gender | Male (including trans male) | 0.511 | 0.459 |
| | Female (including trans female) | 0.486 | 0.536 |
| | Gender queer/Non-binary | 0.004 | 0.004 |
| Race | Caucasian | 0.484 | 0.441 |
| | Latino-Hispanic | 0.084 | 0.094 |
| | Middle Eastern | 0.030 | 0.042 |
| | African | 0.040 | 0.031 |
| | Caribbean | 0.004 | 0.005 |
| | South Asian | 0.128 | 0.145 |
| | East Asian | 0.191 | 0.201 |
| | Pacific Islander | 0.009 | 0.007 |
| | American Indian | 0.002 | 0.002 |
| | Mixed ethnicity | 0.026 | 0.031 |
| Region | Europe | 0.390 | 0.357 |
| | Africa | 0.024 | 0.022 |
| | Asia (including Middle East) | 0.228 | 0.264 |
| | Australasia | 0.035 | 0.031 |
| | North or Central America | 0.290 | 0.289 |
| | South America | 0.034 | 0.037 |
| Caring | No | 0.799 | 0.793 |
| | Yes | 0.201 | 0.207 |
| Job | No | 0.812 | 0.810 |
| | Yes | 0.188 | 0.190 |
| Same country as grown | No | 0.355 | 0.371 |
| | Yes | 0.645 | 0.629 |
| PhD related to original expectation | Meets original expectations | 0.697 | 0.393 |
| | Does not meet original expectations | 0.131 | 0.561 |
| | Exceeds original expectations | 0.171 | 0.046 |
| Hours/week spent on PhD program | Less than 40 h | 0.227 | 0.181 |
| | 41–50 h | 0.300 | 0.259 |
| | more than 50 h | 0.442 | 0.507 |
| | Other, please specify | 0.032 | 0.053 |
| Hours/week contact with supervisor | Less than an hour | 0.443 | 0.610 |
| | Between one and three hours | 0.410 | 0.305 |
| | More than three hours | 0.147 | 0.085 |
| Seek help for anxiety or depression | No | 0.740 | 0.947 |
| | Yes | 0.260 | 0.053 |
| Discrimination | No | 0.873 | 0.545 |
| | Yes | 0.127 | 0.455 |
| Bullying | No | 0.889 | 0.717 |
| | Yes | 0.111 | 0.283 |
| Age: mean (st.dev.) | | 29.41 (6.091) | 30.07 (5.175) |
| N | | 2693 | 3218 |